\documentclass[12pt]{article}

\newcommand{\text}{\textrm}
\usepackage{a4wide}

\def\Red{}
\def\Black{}
\def\Blue{}
\renewcommand{\theequation}{\thesection.\arabic{equation}}
\newcommand{\sect}[1]{ \section{#1} \setcounter{equation}{0} }

\begin{document}

\title{\textbf{{\Red Gluon-ghost condensate of mass dimension 2 in the
Curci-Ferrari gauge }}\Black}
\author{D. Dudal\thanks{%
Research Assistant of The Fund For Scientific Research-Flanders, Belgium.} \
and H. Verschelde\thanks{%
david.dudal@rug.ac.be, henri.verschelde@rug.ac.be} \\
{\small {\textit{Ghent University }}}\\
{\small {\textit{Department of Mathematical Physics and Astronomy,
Krijgslaan 281-S9, }}}\\
{\small {\textit{B-9000 Gent, Belgium}}} \and V.E.R. Lemes, M.S. Sarandy and
S.P. Sorella\thanks{%
vitor@dft.if.uerj.br, sarandy@dft.if.uerj.br, sorella@uerj.br} \\
{\small {\textit{UERJ - Universidade do Estado do Rio de Janeiro,}}} \\
{\small {\textit{\ Rua S\~{a}o Francisco Xavier 524, 20550-013 Maracan\~{a},
}}} {\small {\textit{Rio de Janeiro, Brazil}}} \and M. Picariello\thanks{%
marco.picariello@mi.infn.it } \\
{\small {\textit{Universit\'{a} degli Studi di Milano, via Celoria 16,
I-20133, Milano, Italy }}}\\
{\small {\textit{and INFN\ Milano, Italy}}}}
\maketitle

{\Blue\begin{abstract} The effective potential for an on-shell
BRST invariant gluon-ghost condensate of mass dimension 2 in the
Curci-Ferrari gauge in $SU(N)$ Yang-Mills is analysed by combining
the local composite operator technique with the algebraic
renormalization. We pay attention to the gauge parameter
independence of the vacuum energy obtained in the considered
framework and discuss the Landau gauge as an interesting special
case.
\end{abstract}\Black}

\vfill\newpage\ \makeatother
\renewcommand{\theequation}{\thesection.\arabic{equation}}

\sect{Introduction}

Nowadays an increasing evidence has been reported on the relevance
of the local composite operator $A^{2}$ in the Landau gauge, both
from a phenomenological point of view
\cite{Gubarev:2000eu,Gubarev:2000nz} as from lattice studies
\cite{Boucaud:2002nc,Boucaud:2001st,Burgio:1997hc}. It is no
coincidence that the Landau gauge is used because then $A^{2}$
equals the non-local gauge invariant operator
$(VT)^{-1}\min_{U}\int d^{4}x \left(A^{2}\right)^{U}$ with $VT$
the space time volume. The lattice also revealed that gluons
attain a dynamical mass, see e.g. \cite
{Langfeld:2001cz,Alexandrou:2001fh}. Some older work already
discussed the pairing of gluons in connection with a mass
generation, as a result of the fact that the perturbative
Yang-Mills (YM) vacuum (trivially zero) is unstable
\cite{Fukuda:1977mz,Fukuda:1977zp,Gusynin:1978tr}. More recently,
the connection between a condensate $\left\langle
A^{2}\right\rangle$ and a gluon mass has been made within the OPE
framework \cite{Kondo:2001nq,kmsi}. A technique to effectively
calculate $\left\langle A^{2}\right\rangle$ and the gluon mass was
presented in \cite{v1}, also in the Landau gauge. \newline
\newline
The answer to the question how a mass is generated could be posed in a more
general context than the Landau gauge. The Landau gauge is a limiting case
of a class of renormalizable, generalized covariant gauges introduced in
\cite{Delbourgo:1981cm,Baulieu:sb}. We are therefore led to search for a
local operator which could replace $A^{2}$. A proposal has been made in \cite
{Kondo:2001nq}, where it was shown that $A^{2}$ is a special case of a more
general mass dimension 2 operator, namely $\mathcal{O}=\frac{1}{2}%
A_{\mu}^{a}A^{\mu a}+\alpha \overline{c}^{a}c^{a}$, also involving
ghosts and which is BRST invariant on-shell, however not gauge
invariant (see also \cite{Gripaios:2003xq}). This should allow a
BRST invariant treatment of the mass generation in those gauges.
The proposed condensate is not that surprising, since it equals
the operator coupled to the mass term of a massive, renormalizable
$SU(N)$ model, introduced in \cite {Curci:bt,Curci:1976ar}. The
specific form of the mass term is necessary to maintain the BRST
invariance and renormalizability \cite
{Curci:bt,Curci:1976ar,deBoer:1995dh}. Although the Curci-Ferrari
model (CF) is BRST invariant, the associated BRST operator is not
nilpotent and the model is not unitary
\cite{Ojima:1981fs,deBoer:1995dh}. Since the gauge fixing terms of
the CF model and the YM theories with the gauges discussed in
\cite{Kondo:2001nq,kmsi,Delbourgo:1981cm,Baulieu:sb} are the same,
it seems natural to search in that direction for a suitable
operator that gets a non-vanishing vacuum expectation value and
invokes a dynamical mass.
\newline
\newline
The aim of this paper is to construct an effective potential for the mass
dimension 2 condensate in the CF gauge. It is organized as follows. In
section \ref{sec1} we discuss the formalism to obtain a well-defined
effective potential for the local composite operator $\mathcal{O}=\frac{1}{2}%
A_{\mu}^{a}A^{\mu a}+\alpha \overline{c}^{a}c^{a}$, a non-trivial
task due to the compositeness of this operator \cite{v1,
Knecht:2001cc}. In section \ref{sec2}, we denote the Ward
identities of the action, ensuring the renormalizability. A
further construction of the effective action is discussed in
section \ref{sec3}, where we also outline a subtlety on the
minimization of the effective potential. In section \ref{sec4}, we
consider the gauge parameter independence of the vacuum energy and
spend some words on the BRST charge. Section \ref{sec5} handles
the explicit evaluation of the effective potential. We also
discuss the interesting role of the Landau gauge as a limiting
case of the CF gauge. We pay attention to the similarities between
CF and the Maximal Abelian gauge (MAG). A mass generating
mechanism for the off-diagonal gluons in the MAG very much
resembles that of the CF gauge, and could be seen as some evidence
for Abelian dominance. As usual, conclusions are formulated in the
last section.

\sect{The LCO formalism}

\label{sec1} For a more detailed introduction to the local composite
operator (LCO) formalism and to the algebraic renormalization technique, the
reader is referred to \cite{v1,Knecht:2001cc}, respectively \cite{book}.%
\newline
\newline
Let us begin by giving the expression for the $SU(N)$ Yang-Mills action in
the CF gauge
\begin{eqnarray}
S &=&S_{YM}+S_{GF+FP}=-\frac{1}{4}\int d^{4}xF_{\mu \nu }^{a}F^{a\mu \nu
}+\int d^{4}x\left( b^{a}\partial _{\mu }A^{a\mu }+\frac{\alpha }{2}%
b^{a}b^{a}+\overline{c}^{a}\partial ^{\mu }D_{\mu }^{ab}c^{b}\right.
\nonumber \\
&-&\left. \frac{\alpha }{2}gf^{abc}b^{a}\overline{c}^{b}c^{c}-\frac{\alpha }{%
8}g^{2}f^{abc}f^{cde}\overline{c}^{a}\overline{c}^{b}c^{d}c^{e}\right)
\label{cfa}
\end{eqnarray}
where
\begin{equation}
D_{\mu }^{ab}\equiv \partial _{\mu }\delta ^{ab}+gf^{acb}A_{\mu }^{c}
\label{dudal0}
\end{equation}
is the usual covariant derivative. In order to investigate if
\begin{equation}
\mathcal{O}=\frac{1}{2}A_{\mu }^{a}A^{a\mu }+\alpha \overline{c}^{a}c^{a}
\label{op}
\end{equation}
gets a non-vanishing vacuum expectation value, we introduce a suitable set
of LCO sources \cite{v1, Knecht:2001cc}. In this case this task is
nontrivial. It turns out that in order to introduce the local operator $%
\mathcal{O}$ in the starting action in a BRST invariant way, three external
sources $J,$ $\eta ^{\mu }$ and $\tau ^{\mu }$ are needed, so that
\begin{equation}
S_{\mathrm{LCO}}=\int d^{4}x\left[ J\mathcal{O} +\frac{\xi }{2}J^{2}-\eta
^{\mu }A_{\mu }^{a}c^{a}-\tau ^{\mu }s(A_{\mu }^{a}c^{a})\right]  \label{lco}
\end{equation}
where $\xi $ is the LCO parameter and $s$ denotes the BRST\ operator acting
as
\begin{eqnarray}
sA_{\mu }^{a} &=&-D_{\mu }^{ab}c^{b}  \nonumber \\
sc^{a} &=&\frac{g}{2}f^{abc}c^{b}c^{c}  \nonumber \\
s\overline{c}^{a} &=&b^{a}  \nonumber \\
sb^{a} &=&-Jc^{a}  \nonumber \\
sJ &=&0  \nonumber \\
s\eta ^{\mu } &=&\partial ^{\mu }J  \nonumber \\
s\tau ^{\mu } &=&\eta ^{\mu }  \label{s}
\end{eqnarray}
The parameter $\xi$ has to be introduced since the introduction of the
source term $J\mathcal{O}$ gives rise to novel vacuum energy divergences
proportional to $J^{2}$. These new divergences, related to those of the
connected Green's function $\left\langle \mathcal{O}(x)\mathcal{O}%
(y)\right\rangle_{c}$ for $x\rightarrow y$, are cancelled by a counterterm $%
\delta\xi\frac{J^{2}}{2}$. \newline
\newline
After introduction of the sources, we still have a BRST invariant action
\begin{equation}
s\left( S_{YM}+S_{GF+FP}+S_{\mathrm{LCO}}\right) =0  \label{sinv}
\end{equation}
but it should be observed that, due to the presence of the sources $\left(
J,\eta ^{\mu },\tau ^{\mu }\right) $, the BRST operator is no more
nilpotent, namely
\begin{eqnarray}
s^{2}\Phi &=&0,\;\;\;\Phi =(A,c,J,\eta ^{\mu })  \nonumber \\
s^{2}\overline{c}^{a} &=&-Jc^{a}  \nonumber \\
s^{2}b &=&-J\frac{g}{2}f^{abc}c^{b}c^{c}  \nonumber \\
s^{2}\tau ^{\mu } &=&\partial ^{\mu }J  \label{s2}
\end{eqnarray}
As a consequence, setting
\begin{equation}
s^{2}=\delta _{J}  \label{dj}
\end{equation}
we have
\begin{equation}
\delta _{J}\left( S_{YM}+S_{GF+FP}+S_{\mathrm{LCO}}\right) =0  \label{djinv}
\end{equation}
The operator $\delta _{J}\;$is related to the $SL(2,R)$ symmetry \cite
{Delbourgo:1981cm,Baulieu:sb,Dudal:2002ye} exhibited by the Curci-Ferrari
action. The generators of this $SL(2,R)$ symmetry are, next to the
Faddeev-Popov ghost number $\delta_{FP}$, given by
\begin{eqnarray}
\delta \overline{c}^{a} &=&c^{a}  \nonumber \\
\delta b^{a} &=&\frac{g}{2}f^{abc}c^{b}c^{c}  \nonumber \\
\delta A_{\mu }^{a} &=&\delta c^{a}=0  \label{df}
\end{eqnarray}
and
\begin{eqnarray}
\overline{\delta} c^{a} &=&\overline{c}^{a}  \nonumber \\
\overline{\delta} b^{a} &=&\frac{g}{2}f^{abc}\overline{c}^{b}\overline{c}^{c}
\nonumber \\
\overline{\delta} A_{\mu }^{a} &=&\overline{\delta} \overline{c}^{a}=0
\label{dfbar}
\end{eqnarray}
The action of the $\delta$ symmetry can be enlarged to the sources as $%
\delta J=0$, $\delta\eta_{\mu}=0$ and $\delta \tau_{\mu}=0$. Then it is
obvious from (\ref{s2}) that
\begin{equation}  \label{dudal1}
\delta_{J}=\left.s^{2}\right|_{J=0}-J\delta=-J\delta
\end{equation}
Also, expression $\left( \ref{dj}\right) $ shows that, in the massive case,
the $\delta _{J}$-invariance is a consequence of the modified BRST\
transformations. The lack of nilpotency of the BRST operator together with (%
\ref{dj}) are well known features of the CF gauge in the presence of a mass
term \cite{ds}. Next to the $\delta_{J}$ invariance, the action $%
S_{YM}+S_{GF+FP}+S_{LCO}$ is still invariant under the $NO$ algebra\footnote{%
This algebra is generated by the $SL(2,R)$ and (anti-)BRST transformations $s
$ and $\overline{s}$. It is a trivial task to check that the action is also
anti-BRST invariant, and relations similar to (\ref{dj}), (\ref{djinv}) and (%
\ref{dudal1}) arise for the anti-BRST transformation $\overline{s}$.} \cite
{Dudal:2002ye}, meaning that irrespective of the fact that $\left\langle
\mathcal{O}\right\rangle$ gains a non-trivial value, the $NO$ (and thus the $%
SL(2,R)$) symmetry is unaffected. \newline
\newline
Notice that in the present case the operator $s^{2}\;$always contains the
source $J$ which will be set to zero at the end of the computation.

\sect{Ward identities}

\label{sec2}

Let us now translate the previous invariances into Ward identities. To this
purpose, we introduce external sources $\Omega _{\mu }^{a}\;$and $L^{a}\;$%
coupled to the BRST\ variation of $A_{\mu }^{a}\;$and $c^{a}$

\begin{equation}
S_{ext}=\int d^{4}x\left[ -\Omega ^{a\mu }D_{\mu }^{ab}c^{b}\;+L^{a}\frac{g}{%
2}f^{abc}c^{b}c^{c}\right]  \label{om}
\end{equation}
with

\[
s\Omega _{\mu }^{a}=sL^{a}=0
\]
The complete action
\begin{equation}
\Sigma =S_{YM}+S_{GF+FP}+S_{\mathrm{LCO}}+S_{ext}  \label{sl}
\end{equation}
turns out to obey the following identities:

\begin{itemize}
\item  The Slavnov-Taylor identity
\begin{equation}
\mathcal{S}(\Sigma )=0  \label{st}
\end{equation}
with
\begin{equation}
\mathcal{S}(\Sigma )=\int d^{4}x\left( \frac{\delta \Sigma }{\delta A_{\mu
}^{a}}\frac{\delta \Sigma }{\delta \Omega ^{a\mu }}+\frac{\delta \Sigma }{%
\delta L^{a}}\frac{\delta \Sigma }{\delta c^{a}}+b^{a}\frac{\delta \Sigma }{%
\delta \overline{c}^{a}}+\partial _{\mu }J\frac{\delta \Sigma }{\delta \eta
_{\mu }}+\eta ^{\mu }\frac{\delta \Sigma }{\delta \tau ^{\mu }}-Jc^{a}\frac{%
\delta \Sigma }{\delta b^{a}}\right)   \label{sto}
\end{equation}

The $\delta _{J}$ Ward identity
\begin{equation}
\mathcal{W}(\Sigma )=0  \label{di}
\end{equation}
\end{itemize}

with
\begin{equation}
\mathcal{W}(\Sigma )=\int d^{4}x\left( Jc^{a}\frac{\delta \Sigma }{\delta
\overline{c}^{a}}+J\frac{\delta \Sigma }{\delta L^{a}}\frac{\delta \Sigma }{%
\delta b^{a}}-\partial _{\mu }J\frac{\delta \Sigma }{\delta \tau _{\mu }}%
\right)  \label{dio}
\end{equation}
Proceeding as in \cite{Dudal:2002pq}, these identities imply the
renormalizability of the model and, in particular, the multiplicative
renormalizability of the local operator $\mathcal{O}$.

\sect{Renormalizability of $\mathcal{O}$ and the effective action}

\label{sec3}

As established explicitly in \cite{kmsi,Gracey:2002yt}, the operator $%
\mathcal{O}$ is indeed multiplicative renormalizable in the CF gauge.
Denoting the bare operator by $\mathcal{O}_{B}$, one has
\begin{equation}
\mathcal{O}_{B}=Z_{\mathcal{O}}\mathcal{O}_{R}  \label{mr}
\end{equation}
with\footnote{%
We use dimensional regularization in $d=4-\varepsilon$ dimensions and employ
the $\overline{MS}$ renormalization scheme.} \cite{kmsi,Gracey:2002yt}
\begin{equation}
Z_{\mathcal{O}}=1+\left[ \frac{35}{6}-\frac{\alpha }{2}\right]\frac{g ^{2}N}{%
16\pi ^{2}}\frac{1}{\varepsilon }+\left[\left(\frac{2765}{72}-\frac{11\alpha%
}{3}\right)\frac{1}{\varepsilon^{2}} +\left(\frac{\alpha^{2}}{16}+\frac{%
11\alpha}{16}-\frac{449}{48}\right)\frac{1}{\varepsilon}\right]\left(\frac{%
g^{2}N}{16\pi^{2}}\right)^{2}+\ldots  \label{zo}
\end{equation}
For the anomalous dimension $\gamma _{\mathcal{O}}\;$ of $\mathcal{O}$, one
has \cite{kmsi,Gracey:2002yt}
\begin{equation}
\gamma _{\mathcal{O}}(g^{2},\alpha)=-\mu \frac{\partial \ln Z_{\mathcal{O}}
}{\partial {\mu }}=\left( \frac{35}{6}-\frac{\alpha }{2}\right) \frac{g^{2}N%
}{16\pi ^{2}}+\left(\frac{449}{24}-\frac{\alpha^{2}}{8}-\frac{11\alpha}{8}%
\right)\left(\frac{g^{2}N}{16\pi ^{2}}\right)^{2}+\ldots  \label{go}
\end{equation}
Notice that $\gamma _{\mathcal{O}}$ depends on the gauge parameter $\alpha $%
. This is due to the explicit dependence from $\alpha $ of the operator $%
\mathcal{O}$. Moreover, in the limit $\alpha \rightarrow 0$, expression $%
\left( \ref{go}\right) $ reduces to the anomalous dimension of the Landau
gauge \cite{v1}. Let us also give, for further use, the $\beta $-function of
the gauge parameter $\alpha $ in the CF gauge \cite{kmsi,Gracey:2002yt}.
\begin{equation}
\beta _{\alpha }(g^{2},\alpha)=\frac{\mu}{\alpha}\frac{\partial\alpha}{%
\partial\mu}= \left( \frac{13}{3}-\frac{\alpha }{2}\right) \frac{g^{2}N}{%
16\pi ^{2}}-\frac{\alpha^{2}+17\alpha-118}{16}\left(\frac{g^{2}N}{16\pi ^{2}}%
\right)^2+\ldots  \label{ba}
\end{equation}
In order to obtain the effective potential for the operator $\mathcal{O}$,
we set to zero the sources $\Omega _{\mu }^{a}$, $L^{a},$ $\eta ^{\mu }$ and
$\tau ^{\mu }$, obtaining for the generating functional the following
expression
\begin{equation}
\exp -i\mathcal{W}(J)=\int [D\phi ]\exp iS(J)  \label{wj}
\end{equation}
with
\begin{equation}
S(J)=S_{YM}+S_{GF+FP}+\int d^{4}x\left[ J\mathcal{O} +\frac{\xi }{2}%
J^{2}\right]  \label{sj}
\end{equation}
and $\phi$ denoting the relevant fields.\newline
\newline
From the bare Lagrangian associated to (\ref{sj}), one obtains that the
quantity $\xi(\mu)$ obeys the following renormalization group equation (RGE)
\begin{equation}  \label{d1}
\mu\frac{d\xi}{d\mu}=2\gamma_{\mathcal{O}}(g^{2},\alpha)\xi+\delta(g^{2},%
\alpha)
\end{equation}
where
\begin{equation}  \label{d2}
\delta(g^{2},\alpha)=\left(\varepsilon+2\gamma_{\mathcal{O}%
}(g^{2},\alpha)-\beta(g^{2})\frac{\partial }{\partial g^{2}}%
-\alpha\beta_{\alpha}(g^{2},\alpha)\frac{\partial}{\partial\alpha}%
\right)\delta\xi
\end{equation}
Now, following \cite{v1}, it is possible to set the hitherto \emph{free}
parameter $\xi$ such a function of $g^{2}$ and $\alpha$, so that if $g^{2}$
runs according to $\beta(g^{2})$ and $\alpha$ to $\beta_{\alpha}\left(g^{2}%
\right)$, $\xi(g^{2},\alpha)$ will run according to its RGE (\ref{d1}).
Specifying, $\xi(g^{2},\alpha)$ is the particular solution of
\begin{equation}  \label{d3}
\left(\beta(g^{2})\frac{\partial }{\partial g^{2}}+\alpha\beta_{%
\alpha}(g^{2},\alpha)\frac{\partial}{\partial \alpha}\right)\xi(g^{2},%
\alpha)=2\gamma_{\mathcal{O}}(g^{2},\alpha)\xi(g^{2},\alpha)+\delta(g^{2},%
\alpha)
\end{equation}
Furthermore\footnote{%
The integration constant showing up when (\ref{d3}) is solved, has been put
to zero according to \cite{v1}.}, $\xi(g^{2},\alpha)$ is multiplicatively
renormalizable ($\xi+\delta\xi=Z_{\xi}\xi$). It is easy to see that $%
\xi(g^{2},\alpha)$ will be of the form
\begin{equation}  \label{d4}
\xi(g^{2},\alpha)=\frac{\xi_{0}(\alpha)}{g^{2}}+\xi_{1}(\alpha)+\xi_{2}(%
\alpha)g^{2}+\ldots
\end{equation}
Performing the calculation at 1-loop, we find that
\begin{equation}  \label{d5}
\delta\xi=-\frac{\left(N^{2}-1\right)}{16\pi^{2}}\frac{\left(3-\alpha^{2}%
\right)}{\varepsilon}
\end{equation}
Consequently, solving (\ref{d3}) for $\xi_{0}$ as a function of the gauge
parameter $\alpha$, one finds
\begin{eqnarray}  \label{d6}
\xi_{0}(\alpha)&=&\frac{9}{13}\frac{N^{2}-1}{N}s_{0}(\alpha) \\
s_{0}(\alpha)&=&1+\frac{311}{117}\alpha+6\alpha\left(1-\frac{3\alpha}{26}%
\right)\ln\left|-\frac{26}{\alpha}+3\right|+c\alpha(-26+3\alpha)
\end{eqnarray}
with $c$ an integration constant. Notice that $s_{0}(0)=1$, so that we
recover the result of \cite{v1} in the case of the Landau gauge. In the next
section, we will show that the vacuum energy is gauge parameter independent.
Henceforth, we can forget about the integration constant and set $c=0$.
\newline
\newline
Taking now the functional derivative of $\mathcal{W}(J)$ with respect to $J$%
, we obtain
\begin{equation}
\left. \frac{\delta \mathcal{W}(J)}{\delta J}\right| _{J=0}=-\left\langle%
\mathcal{O}\right\rangle  \label{fd}
\end{equation}
The presence of the $J^{2}$ term in $\mathcal{W}(J)$ seems to spoil an
energy interpretation. However, this can be dealt with by introducing a
Hubbard-Stratonovich field $\sigma$ so that
\begin{equation}
J\mathcal{O}+\frac{\xi }{2}J^{2}\Rightarrow -\frac{\sigma ^{2}}{2\xi g^{2}}+%
\frac{\sigma }{g\xi}\mathcal{O}+\frac{\sigma }{g}J-\frac{1}{2\xi }\mathcal{O}%
^{2}  \label{hs}
\end{equation}
Therefore
\begin{equation}
\exp -i\mathcal{W}(J)=\int [D\phi ]\exp i\left( S_{\sigma }+\int d^{4}x\frac{%
\sigma }{g}J\;\right)  \label{wjs}
\end{equation}
where
\begin{eqnarray}
S_{\sigma } &=&S_{YM}+S_{GF+FP}+\int d^{4}x\left( -\frac{\sigma ^{2}}{2\xi
g^{2}}+\frac{\sigma }{g\xi}\mathcal{O} - \frac{1}{2\xi } \mathcal{O}
^{2}\right)  \label{ss}
\end{eqnarray}
$J$ now appears as a linear source. Hence, we have back an energy
interpretation and the $1PI$ machinery applies. \newline
\newline
Differentiating the functional generator with respect to $J$, one gets the
relationship
\begin{equation}
\left\langle \sigma \right\rangle _{S_{\sigma }}=g\left\langle \mathcal{O}%
\right\rangle  \label{rel}
\end{equation}
Recapitulating, we have constructed a multiplicatively renormalizable action
$S_{\sigma}$ incorporating the effects of a possible non-vanishing vacuum
expectation value for $\mathcal{O}$. The corresponding effective action $%
\Gamma$ obeys a linear, homogeneous RGE. Notice that to get actual knowledge
of the $n$-loop effective action, one needs the values of $\xi_{0}, \ldots,
\xi_{n}$. This means, recalling (\ref{d3}), that we need the $(n+1)$-loop
values of the renormalization group functions. In \cite{Lemes:2002rc}, a
slightly different Hubbard-Stratonovich transformation was used, so that
\begin{equation}
J\mathcal{O}+\frac{\xi }{2}J^{2}\Rightarrow -\frac{\sigma ^{2}}{2 g^{2}}+%
\frac{\sigma }{g\sqrt{\xi}}\mathcal{O}+\frac{\sqrt{\xi}\sigma }{g}J-\frac{1}{%
2\xi }\mathcal{O}^{2}  \label{hsbis}
\end{equation}
resulting in
\begin{equation}
\exp -i\mathcal{W}(J)=\int [D\phi ]\exp i\left( S_{\sigma }+\int d^{4}x\frac{%
\sqrt{\xi}\sigma }{g}J\;\right)  \label{wjsbis}
\end{equation}
where
\begin{eqnarray}
S_{\sigma } &=&S_{YM}+S_{GF+FP}+\int d^{4}x\left( -\frac{\sigma ^{2}}{2g^{2}}%
+\frac{\sigma }{g\sqrt{\xi}}\mathcal{O}-\frac{1}{2\xi }\mathcal{O}
^{2}\right)  \label{ssbis}
\end{eqnarray}
With this action, it seems that it suffices to know $\xi_{0},\ldots,\xi_{n-1}
$ to construct the $n$-loop effective potential. However, some attention
should be paid here. It is indeed so that with (\ref{ssbis}), we do not need
$\xi_{n}$ for $\Gamma_{n-\text{{\tiny {loop}}}}$, but since the source $J$
is now coupled to the operator $\frac{\sqrt{\xi}\sigma}{g}$, we formally
have for the effective action $\Gamma$, being the Legendre transform of $%
\mathcal{W}(J)$
\begin{equation}  \label{dudal2}
\Gamma\left(\frac{\sqrt{\xi}\sigma}{g}\right)=-\mathcal{W}(J)-\int d^{4}y
J(y)\frac{\sqrt{\xi}\sigma(y)}{g}
\end{equation}
Hence
\begin{equation}  \label{dudal3}
\frac{\delta}{\delta\left(\frac{\sqrt{\xi}\sigma(y)}{g}\right)}\Gamma\left(%
\frac{\sqrt{\xi}\sigma(x)}{g}\right)=-J(y)
\end{equation}
Since
\begin{eqnarray}
\Gamma &=& \frac{\Gamma_{0}}{g^{2}}+\Gamma_{1}+\ldots \\
\frac{\sqrt{\xi}}{g} &=& \sqrt{\xi_{0}}\left(\frac{1}{g^{2}}+\frac{\xi_{1}}{%
\xi_{0}}+\ldots\right)
\end{eqnarray}
it becomes clear that, in order to have $J=0$ up to the considered order in
a $g^{2}$ expansion (i.e. to end up in the vacuum state), one must solve
(for constant configurations)
\begin{equation}  \label{dudal4}
\frac{d}{d\left(\frac{\sqrt{\xi}\sigma}{g}\right) }V=0
\end{equation}
which will \emph{not}\footnote{%
Because $\xi$ itself is a series in $g^{2}$.} produce the same (correct) $%
\sigma_{min}$ as by solving
\begin{equation}  \label{dudal5}
\frac{dV}{d \sigma}=0
\end{equation}
as it was done in \cite{Lemes:2002rc}. The most efficient way to
solve (\ref {dudal4}) is by performing the transformation
\begin{equation}  \label{dudal6}
\sigma\rightarrow\frac{\sigma}{\sqrt{\xi}}
\end{equation}
and this exactly transforms the action (\ref{ssbis}) into the one
of (\ref {ss}). Notice that the action (\ref{ssbis}) is not
incorrect, one should only be careful how the vacuum configuration
is constructed. The conclusion is that one cannot escape the job
of doing $(n+1)$-loop calculations for $n $-loop results. \newline
\newline
We draw attention to the fact that the action $S_{\sigma}$ is BRST invariant%
\footnote{%
It is obvious that $s\sigma=gs\mathcal{O}$.}, while this BRST
transformation is nilpotent for $J=0$. This means that the action,
evaluated in its minimum, i.e. the vacuum energy, should be
independent of the gauge parameter $\alpha$ order by order. In the
next section, we pay some more attention to this $\alpha$
independence.

\sect{Gauge parameter independence of the vacuum energy}

\label{sec4} We begin our argumentation from the generating
functional (\ref {wjs}). It will be useful to consider also the
'original' action $\widetilde{S}(J)$ (i.e. before the
Hubbard-Stratonovich
transformation) defined in (\ref{sj}). To avoid confusion with (%
\ref{wjs})-(\ref{ss}), we added a $\sim$ to the notation. The relation
between $\mathcal{W}(J)$ and $\widetilde{S}(J)$ is obtained via the
insertion of a unity
\begin{equation}  \label{19}
1=\frac{1}{N}\int [D\sigma]\exp\left[i\int d^{4}x\left(-\frac{1}{2\xi}\left(%
\frac{\sigma}{g}-\mathcal{O}-\xi J\right)^{2}\right)\right]
\end{equation}
with $N$ an appropriate normalization factor. Explicitly, we have
\begin{equation}  \label{20}
\exp(-i\mathcal{W}(J))=\int[D\phi][D\sigma]\exp i\left[\widetilde{S}(J)+\int
d^{4}x\left(-\frac{1}{2\xi}\left(\frac{\sigma}{g}-\mathcal{O}-\xi
J\right)^{2}\right)\right]
\end{equation}
Since evidently
\begin{equation}  \label{23}
\frac{d}{d\alpha}\frac{1}{N}\int [D\sigma]\exp\left[i\int d^{4}x\left(-\frac{%
1}{2\xi}\left(\frac{\sigma}{g}-\mathcal{O}-\xi J\right)^{2}\right)\right]=0
\end{equation}
we find
\begin{equation}  \label{24}
-\frac{d\mathcal{W}(J)}{d\alpha}=\left\langle s\left(\frac{\overline{c}b}{2}-%
\frac{g^{2}}{4}f^{abc}\overline{c}^{a}\overline{c}^{b}c^{c}\right)\right%
\rangle_{J=0}+\textrm{terms proportional to } J
\end{equation}
The effective action $\Gamma$ is related to $\mathcal{W}(J)$ through a
Legendre transformation
\begin{equation}  \label{25bis}
\Gamma\left(\frac{\sigma}{g}\right)=-\mathcal{W}(J)-\int d^{4}y J(y)\frac{%
\sigma(y)}{g}
\end{equation}
The effective potential $V(\sigma)$ is then defined as
\begin{equation}  \label{25}
-V(\sigma)\int d^{4}x=\Gamma\left(\frac{\sigma}{g}\right)
\end{equation}
Let $\sigma_{min}$ be the solution of
\begin{equation}  \label{26}
\left.\frac{dV(\sigma)}{d\sigma}\right|_{\sigma=\sigma_{min}}=0
\end{equation}
Hence, we have that\footnote{%
To have (\ref{26bis}) correct at any order in $g^{2}$, the minima should be
computed correctly, as explained in the previous section.}
\begin{equation}  \label{26bis}
\sigma=\sigma_{min} \Rightarrow J=0
\end{equation}
Invoking (\ref{26bis}), we derive from (\ref{25bis})-(\ref{25})
\begin{equation}  \label{27}
\left.\frac{d}{d\alpha}V(\sigma)\right|_{\sigma=\sigma_{min}}\int d^{4}x=
\left.\frac{d}{d\alpha}\mathcal{W}(J)\right|_{J=0}
\end{equation}
Finally, combining (\ref{24}) and (\ref{27}), we conclude that
\begin{equation}  \label{28}
\left.\frac{d}{d\alpha}V(\sigma)\right|_{\sigma=\sigma_{min}}=0
\end{equation}
Some extra words concerning (\ref{26bis}) and its consequences (\ref{27})-(%
\ref{28}) are in order. Obviously, this is based on the relation
\begin{equation}  \label{29}
\frac{\delta}{\delta\left(\frac{\sigma}{g}\right)}\Gamma=-J
\end{equation}
An explicit evaluation of the effective potential results in a series for $%
V(\sigma)$, and consequently in a gap equation via (\ref{26}). Said otherwise, $J=0$ means in practice that $J$ equals zero up to a certain order in $%
g^{2}$ as a consequence of the solved gap equation, which is of the form
\begin{equation}  \label{30}
V_{0}(\sigma)+V_{1}(\sigma)g^{2}+\ldots+V_{n-1}(\sigma)\left(g^{2}%
\right)^{n-1}=0
\end{equation}
Returning to (\ref{24}), the terms proportional to $J$ are themselves some
series in $g^{2}$. This means that the product of such a term with $J$ is
again a series, which has to be cut off at the considered order; thus some
terms are dropped. When (\ref{29})-(\ref{30}) are used, it turns out that
the product of such a term with $J$ is also zero, but up to terms of higher
order. Henceforth, the gauge parameter independence is \emph{not exact}, but
holds up to terms of higher order. The same holds true for the BRST charge $%
Q_{BRST}$, which will not be exactly nilpotent, but again up to higher order
terms. As it is well known, $Q_{BRST}$ is used to define physical states as
those annihilated by $Q_{BRST}$ and which are not exact (i.e. $\neq
Q_{BRST}\left|\text{something}\right\rangle$). The nilpotency of $Q_{BRST}$
is needed to move freely in the space of gauge parameter choices. With all
this in mind, the $\alpha$ derivative of the action is reduced to an exact
BRST variation. This is the usual argument used to show that physical
operators, including the vacuum energy, are independent of the choice for
the gauge parameter $\alpha$ \cite{book}. We underline again that here, all
this is not exact, but only valid up to terms of higher order. \newline
\newline
Concluding this section, we have shown that the effective potential,
evaluated at its minimum (i.e. the vacuum energy), is gauge parameter
independent at any order in a loop ($g^{2}$) expansion, at least up to terms
that are of higher order.

\sect{Evaluation of the 1-loop effective potential}

\label{sec5} In order to evaluate the 1-loop effective potential, it is
sufficient to consider only the quadratic terms of $S_{\sigma }$, namely
\begin{equation}
S_{\sigma }^{quad}=\int d^{4}x\left( -\frac{\sigma ^{2}}{2\xi g^{2}}+%
\overline{c}^{a}\Sigma ^{ab}c^{b}+\frac{1}{2}A^{a\mu }\Omega _{\mu \nu
}^{ab}A^{b\nu }\right)   \label{sseff}
\end{equation}
where
\begin{equation}
\Sigma ^{ab}=\delta ^{ab}\left( \partial ^{2}+\frac{\sigma \alpha }{g\xi }%
\right)   \label{sab}
\end{equation}
and
\begin{equation}
\Omega _{\mu \nu }^{ab}=\delta ^{ab}\left[ \left( \partial ^{2}+\frac{\sigma
}{g\xi }\right) g_{\mu \nu }-\left( 1-\frac{1}{\alpha }\right) \partial
_{\mu }\partial _{\nu }\right]   \label{omab}
\end{equation}
To calculate $V$, we use the background formalism with the trivial
background $A_{\mu }=0$. This means that we restrict ourselves to the pure
short-range contributions to $\left\langle \mathcal{O}\right\rangle $. If we
would like to include long-range effects, we could for example use an
instanton background \cite{Boucaud:2002nc}. An asset of considering only
short-range contributions is that one does not have to worry about Gribov
ambiguities, since these short-range contributions are calculated with a
purely perturbative expansion, and perturbation theory is not affected by
Gribov copies, since the considered distances are ''too short'' to make
different gauge copies aware of each other \cite
{Williams:2002nq,Williams:2002dw,Frishman:1978ke,Frishman:gy}.\newline
\newline
For the 1-loop effective potential we get
\begin{equation}
V_{1}(\sigma )=\frac{\sigma ^{2}}{2\xi _{0}}\left( 1-\frac{\xi _{1}}{\xi _{0}%
}g^{2}\right) +i\ln \det \Sigma ^{ab}-\frac{i}{2}\ln \det \Omega _{\mu \nu
}^{ab}  \label{v}
\end{equation}
In $d$ dimensions, it holds that
\begin{eqnarray}
&&\ln \det \delta ^{ab}\left[ g_{\mu \nu }\left( \partial ^{2}+m^{2}\right)
-\left( 1-\frac{1}{\alpha }\right) \partial _{\mu }\partial _{\nu }\right]
\nonumber  \label{e1} \\
&=&\left( N^{2}-1\right) \left[ (d-1)\text{tr}\ln \left( \partial
^{2}+m^{2}\right) +\text{tr}\ln \left( \frac{\partial ^{2}}{\alpha }%
+m^{2}\right) \right]
\end{eqnarray}
Working up to order $\varepsilon ^{0}$ and order $g^{2}$, we find
\begin{eqnarray}
\label{pota}i\ln \det \Sigma ^{ab} &=&i\left( N^{2}-1\right) \int
\frac{d^{d}k}{\left( 2\pi \right) ^{d}}\ln \left(
-k^{2}+\frac{\sigma \alpha }{g\xi }\right)
\nonumber   \\
&=&-\frac{\left( N^{2}-1\right) }{32\pi ^{2}}\left( \frac{g^{2}\sigma
^{2}\alpha ^{2}}{\xi _{0}^{2}}\right) \left( \ln \frac{g\sigma \alpha }{\xi
_{0}\overline{\mu }^{2}}-\frac{3}{2}-\frac{2}{\varepsilon }\right)  \\
\label{potb}-\frac{i}{2}\ln \det \Omega _{\mu \nu }^{ab}
&=&-\frac{i}{2}\left( N^{2}-1\right) \int \frac{d^{d}k}{\left(
2\pi \right) ^{d}}\left[ (d-1)\ln
\left( -k^{2}+\frac{\sigma }{g\xi }\right) \right.   \nonumber \\
&+&\left. \ln \left( -\frac{k^{2}}{\alpha }+\frac{\sigma }{g\xi }\right)
\right]   \nonumber \\
&=&\frac{3\left( N^{2}-1\right) }{64\pi ^{2}}\left( \frac{g^{2}\sigma ^{2}}{%
\xi _{0}^{2}}\right) \left( \ln \frac{g\sigma }{\xi _{0}\overline{\mu }^{2}}-%
\frac{5}{6}-\frac{2}{\varepsilon }\right)   \nonumber \\
&+&\frac{\left( N^{2}-1\right) }{64\pi ^{2}}\left( \frac{g^{2}\alpha
^{2}\sigma ^{2}}{\xi _{0}^{2}}\right) \left( \ln \frac{g\alpha \sigma \ }{%
\xi _{0}\overline{\mu }^{2}}-\frac{3}{2}-\frac{2}{\varepsilon }\right)
\end{eqnarray}
Subsequently, we obtain for the one-loop effective potential in the $%
\overline{MS}$ scheme\footnote{%
It is easily checked that using the renormalized version of the
Hubbard-Stratonovich transformation (\ref{hs}), the counterterm
proportional to $\delta\xi$ removes the infinities coming from
(\ref{pota}) and (\ref{potb}).}
\begin{eqnarray}
V_{1}(\sigma ) &=&\frac{\sigma ^{2}}{2\xi _{0}}\left( 1-\frac{\xi _{1}}{\xi
_{0}}g^{2}\right) +\frac{3\left( N^{2}-1\right) }{64\pi ^{2}}\left( \frac{%
g^{2}\sigma ^{2}}{\xi _{0}^{2}}\right) \left( \ln \frac{g\sigma }{\xi _{0}%
\overline{\mu }^{2}}-\frac{5}{6}\right)   \nonumber  \label{f1} \\
&-&\frac{\left( N^{2}-1\right) }{64\pi ^{2}}\left( \frac{g^{2}\sigma
^{2}\alpha ^{2}}{\xi _{0}^{2}}\right) \left( \ln \frac{g\alpha \sigma }{\xi
_{0}\overline{\mu }^{2}}-\frac{3}{2}\right)
\end{eqnarray}
with $\xi _{0}$ given by (\ref{d6}). In principle, as soon one knows the
value of $\xi _{1}$, one can set $\overline{\mu }^{2}=\frac{\sigma }{\sqrt{%
\xi _{0}}}$ and use the renormalization group equation for $V(\sigma )$ to
sum leading logarithms and solve the gap equation. This leads to a value for
the vacuum energy $E$, gluon mass $m_{\text{{\tiny {gluon}}}}$, and through
the trace anomaly, one also finds an estimation for $\left\langle \frac{%
\alpha _{s}}{\pi }F^{2}\right\rangle =-\frac{32}{11}E$. Since the
aim of this paper is merely to describe the mass generation
\emph{mechanism} in the CF gauge, we do not perform the 2-loop
calculation leading to $\xi _{1}$ and corresponding numerical
values. Moreover, since the vacuum energy is gauge parameter
independent, we may choose a specific $\alpha $. Therefore, we
restrict ourselves to the case $\alpha =0$, for which $\xi _{1}$
has already been determined \cite{v1}.\newline
\newline
The Landau gauge is by far the most interesting choice. It is a
fixed point of the renormalization group for the gauge parameter
at any order. Due to the transversality condition $\partial _{\mu
}A^{\mu }=0$, it is a quite physical gauge. It has some
interesting non-renormalization properties \cite {book}. Even more
interesting is the already mentioned fact that $\mathcal{O} $
reduces to $A^{2}$, which has a gauge-invariant meaning in the
Landau gauge, since it equals $(VT)^{-1}\min_{U}\int d^{4}x\left(
A^{2}\right) ^{U}$, a gauge-invariant (however in general non-local) operator%
\footnote{%
Although this correspondence is somewhat troubled by Gribov copies \cite
{Stodolsky:2002st}, but this is of no relevance in the presented approach.}.
As a consequence, the gauge invariance\footnote{%
Which is in fact a stronger statement than gauge parameter
independence.} of the formalism is more obvious in the Landau
gauge \cite{v1}. The relevance of the Landau gauge has also been
pointed out from a more topological point of view
\cite{Gubarev:2000nz}. In case of compact 3-dimensional QED,
$A^{2}$ was shown to be an order parameter for the monopole
condensation \cite {Gubarev:2000eu,Gubarev:2000nz}. If monopole
condensation has something to do with confinement, there might
exist a relation between $A^{2}$ and confinement in case of QCD
too. All these things are less clear in the case of the
$\mathcal{O}$ operator in the CF gauge.\newline
\newline
Having said all this, it might look like that our efforts are not
that important for $\alpha \neq 0$. This is however not the case.
We have given a consistent framework to \emph{calculate} the
dynamically generated gluon mass for the CF gauge. Notice that the
obtained Lagrangian in the condensed vacuum is however not the one
of the Curci-Ferrari model \cite {Curci:bt,Curci:1976ar}. The
question, also posed in \cite{v1}, is if the dynamically massive
YM action (\ref{ss}) breaks unitarity? From a pragmatic point of
view, a possible lack of unitarity in the gluon sector should not
be considered very problematic. After all, since gluons are not
observables due to confinement, massive gluons are a fortiori
unphysical. In fact, a deep connection might exist between massive
gluons and confinement, as it was explored in \cite{Kugo:gm}. See
\cite{Kondo:2002xn} for an attempt to construct a string theory
incorporating a $\left\langle A^{2}\right\rangle $ condensate.
\newline
\newline
We notice that the action (\ref{cfa}) can be rewritten as
\begin{equation}
S=S_{YM}+s\overline{s}\int d^{4}x\left( \frac{1}{2}A_{\mu }^{a}A^{\mu a}-%
\frac{\alpha }{2}c^{a}\overline{c}^{a}\right)   \label{cfa2}
\end{equation}
with\footnote{%
We disregard $S_{LCO}$ here.}
\begin{eqnarray}
\overline{s}A_{\mu }^{a} &=&-D_{\mu }^{ab}\overline{c}^{b}  \nonumber \\
\overline{s}\overline{c}^{a} &=&\frac{g}{2}f^{abc}\overline{c}^{b}\overline{c%
}^{c}  \nonumber \\
\overline{s}c^{a} &=&-b^{a}+gf^{abc}c^{b}\overline{c}^{c}  \nonumber \\
\overline{s}b^{a} &=&-gf^{abc}b^{b}\overline{c}^{c}  \label{sbar}
\end{eqnarray}
Another very interesting renormalizable gauge is the modified
Maximal Abelian gauge (MAG) \cite{Kondo:1997pc}, particularly
useful in the context of the dual superconductivity mechanism for
confinement. This gauge partially fixes the local $SU(N)$ freedom,
i.e. up to the Abelian degrees of freedom. The MAG shares a close
similarity with the CF gauge, since its gauge fixing is given by
\begin{equation}
S=S_{YM}+s\overline{s}\int d^{4}x\left( \frac{1}{2}A_{\mu }^{a^{\prime
}}A^{\mu a^{\prime }}-\frac{\alpha }{2}c^{a^{\prime }}\overline{c}%
^{a^{\prime }}\right)   \label{dudal30}
\end{equation}
where the accent means that the color index runs strictly over the
non-Abelian degrees of freedom. In particular, in \cite
{Dudal:2002ye} it has been shown that the remaining Abelian
degrees of freedom can be fixed so that the resulting theory
displays a global $SL(2,R)$ symmetry, in complete analogy with the
CF\ gauge. Furthermore, due to the similarity
(\ref{cfa2})-(\ref{dudal30}), it is not difficult to show that a
quite analogous treatment with a source $J$ coupled to the
$U(1)^{N-1}$ invariant operator
\begin{equation}
\mathcal{O}^{\prime }=\frac{1}{2}A_{\mu }^{a^{\prime }}A^{\mu a^{\prime
}}+\alpha \overline{c}^{a^{\prime }}c^{a^{\prime }}  \label{dudal31}
\end{equation}
will provide us with a dynamical mass for the off-diagonal gluons
and ghosts  \cite
{Kondo:2001nq,Dudal:2002ye,Dudal:2002xe,Ellwanger:2002sj}, a hint
for some kind of Abelian dominance \cite{Kondo:2000ey}. This
strategy for the MAG was already put forward in
\cite{Kondo:2001nq}. Just as the operator $\mathcal{O}$ is
multiplicatively renormalizable in the CF gauge, the operator
$\mathcal{O}^{\prime}$ will be multiplicatively renormalizable in
the MAG \cite{Ellwanger:2002sj}. So far for the similarities
between CF and MAG. Although it would be nice to stretch the
similarity further and simply put $\alpha =0$ from the beginning,
in which case the MAG reads in differential form $D_{\mu
}^{a^{\prime }b^{\prime }}A^{\mu b^{\prime }}=0$ with $D_{\mu
}^{a^{\prime }b^{\prime }}$ the $U(1)^{N-1}$ Abelian covariant
derivative. As such, we have some kind of $U(1)^{N-1}$ invariant
version of the Landau gauge. Unfortunately, the limit $\alpha
\rightarrow 0$ is now far from being trivial
\cite{Schaden:1999ew}. Moreover, $\alpha =0$ is not a fixed point
of the renormalization group \cite
{Schaden:1999ew,Shinohara:2001cw}.  Also, although for $\alpha =0$
the tree level action (\ref{dudal30}) does not contain a 4-ghost
interaction, radiative corrections will reintroduce this
interaction \cite{Kondo:1997pc}, unlike the Landau gauge. Making a
long story short, we are forced to let the gauge parameter $\alpha
$ free and perform a similar analysis as done in the previous
sections. At the
end of such a more general analysis, one could investigate if the limit $%
\alpha \rightarrow 0$ can be taken. \newline
\newline
Before we formulate our conclusion, we quote the results obtained for the
Landau gauge in \cite{v1}
\begin{eqnarray}
\xi _{1} &=&\frac{161}{52}\frac{N^{2}-1}{16\pi ^{2}}  \nonumber \\
\left. \frac{g^{2}N}{16\pi ^{2}}\right| _{1{\tiny {\text{-loop}}}} &=&\frac{%
36}{187}  \nonumber \\
m_{\text{{\tiny {gluon}}}} &\approx &485\textrm{MeV for }N=3
\nonumber
\\
E &\approx &-0.001\textrm{GeV}^{4}\textrm{ for }N=3  \nonumber \\
\left\langle \frac{\alpha _{s}}{\pi }F^{2}\right\rangle  &\approx &0.003%
\textrm{GeV}^{4}\textrm{ for }N=3
\end{eqnarray}
As the relevant expansion parameter, i.e. $g^{2}N/16\pi ^{2}$, is
relatively small and results do not change much if the second loop
correction to $V(\sigma )$ is included \cite{v1}, qualitatively
acceptable
results are achieved. The value for the 1-loop dynamical gluon mass $m_{%
\text{{\tiny {gluon}}}}$ is also in qualitative agreement with lattice
values \cite{Langfeld:2001cz,Alexandrou:2001fh}, reporting something like $%
m_{\text{{\tiny {gluon}}}}\sim 600$ MeV.

\sect{Conclusion}

\label{sec6} In this paper, we have constructed a renormalizable
effective potential for the on-shell BRST invariant local
composite operator of mass dimension 2 in the Curci-Ferrari gauge,
namely $\mathcal{O=}\frac{1}{2}A_{\mu }^{a}A^{\mu a}+\alpha
\overline{c}^{a}c^{a}$. This gauge reduces to the Landau gauge in
the limit $\alpha =0$. It is worth underlining that, in the Landau
gauge, the operator $\mathcal{O}$ equals the gauge invariant
operator $A^{2}$. Much attention has been paid recently to the
condensate $\left\langle A^{2}\right\rangle $. The generalization
to $\alpha \neq 0$ has also its importance due to the close
analogy with the Maximal Abelian gauge, where the $\alpha
\rightarrow 0$ limit is not as obvious as in case of the CF gauge.
In particular, we have shown that the vacuum energy obtained in
the presented formalism for the CF gauge is independent from the
gauge parameter $\alpha $. As already underlined the $\alpha
$-independence has to be understood in a $g^{2}$ expansion and up
to terms of higher order.\newline
\newline
We restricted ourselves in this paper to the on-shell BRST
invariant condensate resulting in a mass for the particles. A
gluon mass modifies the behaviour of the gluon propagator in the
infrared (see e.g. \cite{Langfeld:2001cz}) and might be relevant
for the confinement problem. A more intensive study would also
include the pure ghost condensates, also of mass dimension 2,
discussed in \cite
{Dudal:2002ye,Lemes:2002rc,Dudal:2002xe,Kondo:2000ey,Schaden:1999ew,Lemes:2002ey,Lemes:2002jv}.
These are not directly related to the mass generation for the
gluons \cite{Dudal:2002ye,Dudal:2002xe}, but are relevant for the
$SL(2,R)$ symmetry and can modify the ghost propagator.

\section*{Acknowledgments}
One of us (M.P.) would like to thank R. Ferrari for useful
discussions. The Conselho Nacional de Desenvolvimento
Cient\'{i}fico e Tecnol\'{o}gico CNPq-Brazil, the
Funda{\c{c}}{\~{a}}o de Amparo a Pesquisa do Estado do Rio de
Janeiro (Faperj), the SR2-UERJ and the Ministero dell'Istruzione
dell'Universit\'a e della Ricerca - Italy are acknowledged for the
financial support.

\end{document}